%% file: main.tex
\pdfoutput=1
\documentclass[11pt]{article}

\usepackage[]{acl}
\usepackage{times}
\usepackage{latexsym}
\usepackage{textcomp}
\usepackage[T1]{fontenc}
\usepackage[utf8]{inputenc}
\usepackage{microtype}
\usepackage{inconsolata}
\usepackage{booktabs}
\usepackage{xspace}
\usepackage{amsmath}
\usepackage{amssymb}
\usepackage{mathtools}
\usepackage{tcolorbox}
\usepackage{xspace}
\usepackage{caption}
\usepackage{subcaption}
\usepackage{balance}
\usepackage[hang,flushmargin]{footmisc}

\newcommand{\RepLLaMA}{Rep\-LLaMA\xspace}
\newcommand{\RankLLaMA}{Rank\-LLaMA\xspace}

\title{Fine-Tuning LLaMA for Multi-Stage Text Retrieval}

\author{Xueguang Ma~$^\dag$ \quad  Liang Wang~$^\ddag$ \quad Nan Yang~$^\ddag$ \quad Furu Wei~$^\ddag$\quad Jimmy Lin~$^\dag$\\[1ex]
$^\dag$ David R.\ Cheriton School of Computer Science, University of Waterloo \\
$^\ddag$ Microsoft Research \\[1ex]
}

\begin{document}
\maketitle
\begin{abstract}
The effectiveness of multi-stage text retrieval has been solidly demonstrated since before the era of pre-trained language models.
However, most existing studies utilize models that predate recent advances in large language models (LLMs).
This study seeks to explore potential improvements that state-of-the-art LLMs can bring.
We conduct a comprehensive study, fine-tuning the latest LLaMA model both as a dense retriever (\RepLLaMA) and as a pointwise reranker (\RankLLaMA) for both passage retrieval and document retrieval using the MS MARCO datasets.
Our findings demonstrate that the effectiveness of large language models indeed surpasses that of smaller models.
Additionally, since LLMs can inherently handle longer contexts, they can represent entire documents holistically, obviating the need for traditional segmenting and pooling strategies.
Furthermore, evaluations on BEIR demonstrate that our \RepLLaMA--\RankLLaMA pipeline exhibits strong zero-shot effectiveness.
Model checkpoints from this study are available on HuggingFace.\footnote{\url{https://huggingface.co/castorini}}
\end{abstract}

\section{Introduction}

Text retrieval, which entails identifying and ranking the most relevant documents or text snippets in response to a query, is crucial in various open-domain language comprehension tasks~\cite{petroni-etal-2021-kilt}, including web search~\cite{msmarco}, open-domain question answering~\cite{chen-etal-2017-reading}, and fact verification~\cite{thorne-etal-2018-fever}.
Retrieval also plays an important role in enhancing the effectiveness of large language models (LLMs) in a retrieval-augmented generation (RAG) pipeline~\cite{rag, Shi2023REPLUGRB}.
This approach not only mitigates hallucinations but also enables LLMs to access knowledge that is not captured within their parameters~\cite{Yang2023InferenceWR, jiang2023flare}.

A typical multi-stage text retrieval pipeline consists of a \textit{retriever}, designed to efficiently locate the top-$k$ relevant texts from a corpus, and a \textit{reranker}, which further refines the order of the retrieved candidates to improve output quality~\cite{monobert}.
Both retrievers and rerankers have significantly benefited from the advent of pre-trained language models based on Transformers~\cite{transformers} such as BERT~\cite{bert} and T5~\cite{T5}.
These models are trained to encode queries and documents into vector representations for retrieval~\cite{karpukhin-etal-2020-dense,Lin_arXiv2021_repir} or to directly score the relevance between a query and a document for reranking~\cite{Nogueira2019MultiStageDR, rankT5}.

Recent large language models with billions of parameters, fine-tuned to follow instructions, such as Instruct\-GPT~\cite{instructGPT}, GPT-4~\cite{openai2023gpt4}, and LLaMA~\cite{llama1, llama2}, have exhibited extraordinary capabilities in many NLP tasks, surpassing previous smaller pre-trained language models~\cite{llm_survey}.
For retrieval, recent methods such as LRL~\cite{LRL}, RankGPT~\cite{RankGPT}, and PRP~\cite{PRP} have explored prompting LLMs to perform zero-shot reranking using pairwise or listwise approaches.
These methods leverage LLMs by viewing reranking as text generation.

However, we see a number of potential issues.
First, these methods do not address the entire multi-stage pipeline, as it is challenging to cast retrieval from a large corpus as a text generation task.
Second, they do not leverage labeled data when available.
Finally, these rerankers are not efficient because they do not support parallel scoring and are slowed by their multi-pass decoding design.

Therefore, we argue that fine-tuning state-of-the-art large language models to function as retrievers and rerankers can yield better effectiveness than previous smaller models.
This approach can also optimally utilize LLMs within multi-stage pipelines.
Thus, we are motivated to investigate the following research question: How do state-of-the-art large language models perform when specifically fine-tuned for multi-stage text retrieval?

Our study aims to answer this question by conducting a comprehensive investigation into fine-tuning the latest LLaMA-2 model~\cite{llama2}, a state-of-the-art, open-source large language model, as both a retriever and a reranker, which we refer to as \RepLLaMA and \RankLLaMA, respectively.
Specifically, we utilize the MS MARCO~\cite{msmarco} and BEIR~\cite{beir} datasets for our experiments.
Our findings suggest that large language models surpass previous smaller models, achieving state-of-the-art effectiveness for both retrieval and reranking through a straightforward training regime and exhibiting strong zero-shot effectiveness.
Furthermore, we observe that LLMs, which are inherently pre-trained on longer contexts, demonstrate potential in representing entire documents, thereby eliminating the need for traditional segmenting and pooling strategies for document retrieval.

\section{Method}

\subsection{Preliminaries}

\paragraph{Task Definition} Given a query $Q$ and a corpus $C$ = $\{D_1, D_2, ..., D_n\}$ consisting of $n$ documents, the goal of text retrieval is to find the $k$ documents that are most relevant to the query $Q$, with $k\ll n$.
In a multi-stage retrieval pipeline composed by a retriever and a reranker, the retriever's task is to efficiently generate the top-$k$ candidates that are relevant to the query based on the similarity metric $\textrm{Sim}(Q, D)\in \mathbb{R}$.
The reranker's task is to reorder these $k$ candidate documents further to improve the relevance order using a more effective, but typically more computationally expensive reranking model.
Note that ``document'' in this context can refer to an arbitrary information snippet, including sentences, passages, or full documents.
While a multi-stage pipeline can contain multiple rerankers, in this paper we focus on a single reranker.

Modern retrievers typically follow a bi-encoder architecture that encodes text into vector representations, with $\textrm{Sim}(Q, D)$ computed as the dot product of the vector representations of the query $Q$ and a document $D$~\cite{karpukhin-etal-2020-dense}.
In contrast, a (pointwise) reranker typically takes both the query and a candidate document as input to directly generate a relevance score.
These scores are then used to reorder the candidates~\cite{Nogueira2019MultiStageDR, Gao2021RethinkTO}.

\paragraph{LLaMA} LLaMA~\cite{llama1} is an auto-regressive, decoder-only large language model based on the Transformer architecture.
The model is characterized by its billions of parameters, pre-trained on a vast amount of web data.
Being uni-directional means that the model's attention mechanism only considers the preceding elements in the input sequence when making predictions.
Specifically, given an input sequence \(x = [t_1, t_2, ..., t_{n-1}]\), the model computes the probability of the next token \(t_n\) based solely on the preceding tokens.
The prediction process can be mathematically represented as $P(t_n |t_1, ..., t_{n-1})$, where \( P \) denotes the probability and \( t_n \) represents the next element in the sequence.

\subsection{Retriever}
\label{sec:retriever}

Our retriever model, called \RepLLaMA, follows the bi-encoder dense retriever architecture proposed in DPR~\cite{karpukhin-etal-2020-dense}, but with the backbone model initialized with LLaMA.

Previous work on dense retriever models often uses a bi-directional encoder-only model like BERT, taking the representation of the prepended \texttt{[CLS]} token as the dense representation of the text input.
However, as LLaMA is uni-directional, we append an end-of-sequence token \texttt{</s>} to the input query or document to form the input sequence to LLaMA.
Thus, the vector embedding of a query or a document is computed as:
\begin{equation*}
V_T = \text{Decoder}(\text{`}t_{1}\text{\textvisiblespace}t_{2}\text{\textvisiblespace}...\text{\textvisiblespace}t_{k}\texttt{</s>}\text{'})[-1]
\end{equation*}
\noindent where Decoder($\cdot$) represents the LLaMA model, which returns the last layer token representation for each input token.
We take the representation of the end-of-sequence token as the representation of the input sequence $t_1 \ldots t_k$, which can be either a query $Q$ or a document $D$.
Relevance of $D$ to $Q$ is computed in terms of the dot product of their corresponding dense representation $V_Q$ and $V_D$ as $\text{Sim}(Q, D) = <V_Q, V_D>$.

The model is then optimized end-to-end according to InfoNCE loss:
\begin{equation}\label{eq:dpr_obj}
\centering
\small
\begin{split}
    &\mathcal{L}(Q, D^+, \{D_{\text{N}}\})\nonumber = -\log p(D=D^+\mid Q)\nonumber = \\
    &-\log \frac{\exp(\text{Sim}(Q, D^+))}{\exp( \text{Sim}(Q, D^+)) + \sum\limits_{D_i^-\in\{D_{\text{N}}\}} \exp(\text{Sim}(Q, D_i^-))}
\end{split}
\end{equation}
\noindent Here, $D^+$ represents a document that is relevant to the query $Q$ (based on human labels), while $\{D_N\}$ denotes a set of documents that is not relevant to the query.
The set of negative documents includes both hard negatives, which are sampled from the top-ranking results of an existing retrieval system, and in-batch negatives, which are derived from the positive documents and hard negative documents associated with other queries in the same training batch.
In practice, dense retrieval training tends to benefit from a larger set of hard negatives and in-batch negatives.

During the inference phase, the query is typically encoded in real-time and the top-$k$ similar documents are searched within the pre-encoded corpus using an efficient approximate nearest neighbour search library such as HNSW~\cite{HNSW}.
However, in this study, we opt to perform exact nearest neighbour search using flat indexes to evaluate model effectiveness.

\subsection{Reranker}

Our reranker model, referred to as \RankLLaMA, is trained as a pointwise reranker.
This approach involves passing a query and a candidate document together as model input, with the model generating a score that indicates the relevance of the document to the query~\cite{Nogueira2019MultiStageDR}.

In more detail, \RankLLaMA reranks a query--document pair as follows:
\begin{equation*}
\begin{aligned}
\text{input} = \text{`query: \{$Q$\} document: \{$D$\}\texttt{</s>}'} \\
\textrm{Sim}(Q,D) = \text{Linear}(\text{Decoder}(\text{input})[-1])
\end{aligned}
\end{equation*}
\noindent where Linear($\cdot$) is a linear projection layer that projects the last layer representation of the end-of-sequence token to a scalar.
Similar to the retriever, the model is optimized by contrastive loss.
However, in this case, the negative documents do not involve in-batch negatives.

To train a reranker that is optimized to rerank candidates from a specific retriever in a multi-stage pipeline, hard negatives should be sampled from the top-ranking results from that retriever.
Specifically, in our case, the hard negative training data for \RankLLaMA are selected from the top-ranking results of \RepLLaMA.

During the inference stage, the top candidate documents retrieved by \RepLLaMA are reordered.
This reordering is based on the relevance score that \RankLLaMA assigns to each query--document pair, with the documents arranged in descending order of relevance.

\input{results_passage}

\section{Experiments}
\label{section:experiments}

We conduct experiments on MS MARCO passage ranking and document ranking datasets to investigate the effectiveness of the multi-stage text retrieval pipeline built using \RepLLaMA and \RankLLaMA for both passage and document retrieval.

\subsection{Passage Retrieval}

\paragraph{Dataset}
We train our retriever and reranker models with LLaMA on the training split of the MS MARCO passage ranking dataset~\cite{msmarco}, which consists of approximately 500k training examples.
As discussed in Section~\ref{sec:retriever}, the incorporation of hard negatives is crucial for the effective training of the retriever.
In our case, we use a blend of BM25 and CoCondenser~\cite{gao-callan-2022-unsupervised} hard negatives to ensure that the hard negatives are derived from both sparse and dense retrieval results, thereby enhancing the diversity of the samples.
For the reranker, we select the hard negatives from the top-$200$ candidates generated by the retriever.

We evaluate the effectiveness of our models using the development split of the MS MARCO passage ranking task, comprising 6980 queries.
Effectiveness is reported using MRR@10 as the metric.
In addition, we also evaluate our models on the TREC DL19/DL20 passage ranking test collections~\cite{dl19, dl20}, which include 43 and 54 queries, respectively.
These collections utilize the same passage corpus as MS MARCO, but provide query sets with dense, graded human relevance judgments.
Following standard practice, we adopt nDCG@10 as the evaluation metric in our experiments.

In addition, we assess the zero-shot effectiveness of \RepLLaMA and \RankLLaMA on BEIR~\cite{beir}, which is a compilation of 18 datasets that spans a variety of domains (e.g., news, medical) and retrieval tasks (e.g., fact verification, question answering).
We focus our evaluation on the 13 datasets that are publicly available.

\paragraph{Implementation Details}
We initialize our models with the LLaMA-2-7B checkpoint\footnote{\url{https://huggingface.co/meta-llama/Llama-2-7b-hf}} and train on 16 $\times$ 32G V100 GPUs.
For \RepLLaMA, we extract the final layer representation of the \texttt{</s>} token as the dense representation, which has a dimensionality of 4096.
Additionally, we normalize these dense representations into unit vectors during both the training and inference stages, ensuring that their L2-norms are equal to 1.
After pre-encoding the entire corpus, we end up with a 135G flat index for brute-force search.

A challenge in fine-tuning LLMs for retrieval is the high GPU memory costs associated with contrastive learning, as it requires large batch sizes for in-batch negatives.
To address this, we employ recent memory efficiency solutions, including LoRA~\cite{hu2022lora}, flash attention~\cite{dao2023flashattention2}, and gradient checkpointing to reduce GPU memory usage.
Both the retriever and reranker are trained with a batch size of 128, with 15 hard negative passages sampled for each query.
At inference time, \RepLLaMA retrieves the top-$1000$ passages from the corpus and \RankLLaMA reranks the top-$200$ passages retrieved by \RepLLaMA.
To explore whether increases in model size can further improve effectiveness, we also train a version of \RankLLaMA using LLaMA-2-13B initialization.\footnote{\url{https://huggingface.co/meta-llama/Llama-2-13b-hf}}

\input{results_beir}

\paragraph{In-Domain Evaluation}
Table~\ref{tab:passage} presents the effectiveness of \RepLLaMA and \RankLLaMA on the MS MARCO passage corpus in comparison to existing methods.

For retrieval, \RepLLaMA outperforms all competing methods, achieving the highest effectiveness.
The closest system in terms of effectiveness is bi-SimLM~\cite{wang-etal-2023-simlm}, which \RepLLaMA outperforms by 2 points MRR@10 on the dev queries.
However, bi-SimLM involves a pre-training stage for enhancing the text representation.
In contrast, \RankLLaMA directly uses the off-the-shelf LLaMA model as initialization.
When compared to the GTR-XXL retriever, which also uses a model with billions of parameters based on the T5-encoder~\cite{ni-etal-2022-large}, our model achieves higher MRR@10 and Recall@1k on the dev queries and on TREC DL19/DL20.
Specifically, \RepLLaMA achieves 2.4 points higher MRR@10 and 0.4 points higher Recall@1k than GTR-XXL.

It is worth noting that recent studies have shown the potential to further improve dense retrieval models by learning from soft labels provided by a reranker via optimizing KL-divergence.
However, in this study, we train our model with only binary judgments.
Training \RepLLaMA by knowledge distillation will likely lead to further improvements, but we save this for future work.

For reranking, \RankLLaMA reranks the top-$200$ passages from \RepLLaMA, resulting in the highest end-to-end effectiveness of any multi-stage retrieval system that we are aware of.
Our complete \RepLLaMA--\RankLLaMA pipeline beats the previous state-of-the-art reranker, RankT5~\cite{rankT5}, by 1.5 points MRR@10.
Furthermore, our \RankLLaMA-13B model outperforms the 7B model, achieving 0.3 points higher MRR@10 and slightly higher nDCG@10 on both DL19 and DL20.
This indicates the potential for further improvements with even larger models.

Compared to RankGPT$_4$~\cite{RankGPT}, which prompts GPT-4 to perform passage reranking through permutation generation within a multi-stage retrieval pipeline, our \RepLLaMA--\RankLLaMA pipeline outperforms it by 0.4 and 7.3 nDCG@10 points on DL19 and DL20, respectively.
As a pointwise reranker, \RankLLaMA can rerank candidate passages in parallel, which means that inference can be accelerated to reduce latency as compared to RankGPT, which depends on a sequential sliding-window strategy to rerank.

\paragraph{Zero-Shot Evaluation}
The zero-shot evaluation of \RepLLaMA and \RankLLaMA on the BEIR datasets is presented in Table~\ref{tab:beir}.
Both models demonstrate superior zero-shot effectiveness, outperforming existing models.
\RepLLaMA surpasses other existing dense retrievers with billions of parameters.
Specifically, it outperforms SGPT~\cite{muennighoff2022sgpt} and Ada2 by 3 points and exceeds GTR-XXL by approximately 6 points.
Note that these methods require an unsupervised contrastive pre-training stage before the supervised fine-tuning.
In contrast, \RepLLaMA uses the base pre-trained model as initialization, achieving the highest zero-shot effectiveness we are aware of while maintaining simplicity.
\RankLLaMA-7B further enhances the retriever's effectiveness by an average of 1.5 points on nDCG@10.
Interestingly, the larger \RankLLaMA-13B model does not appear to yield any further improvements.

\begin{figure*}[t]
\centering
\includegraphics[width=\textwidth]{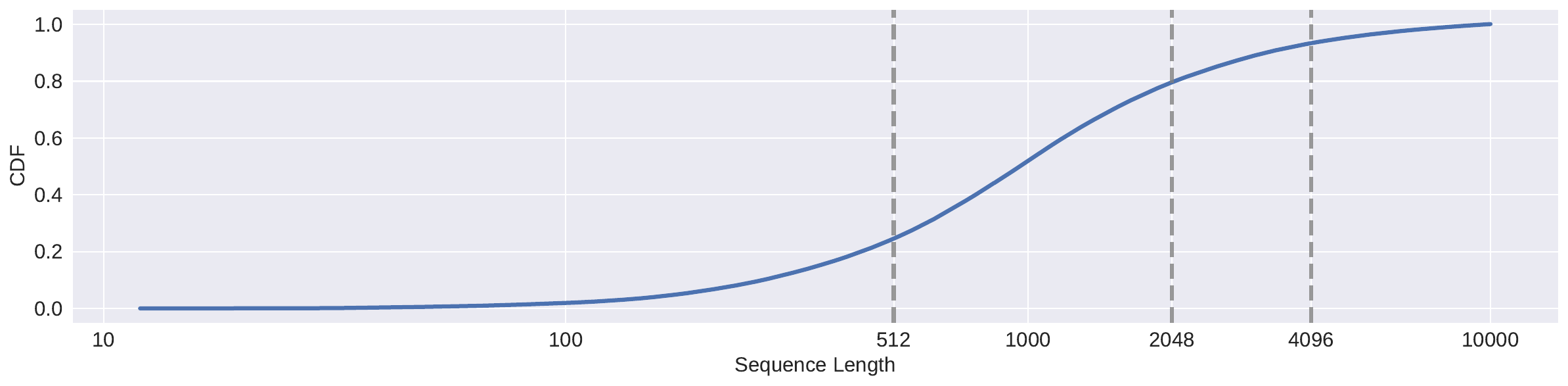}
\caption{Cumulative distribution function of document lengths in the MS MARCO document corpus, showing the proportion of documents that has a length less than a specific value (determined by the LLaMA tokenizer).
For clarity, we exclude 3\% of documents with a length exceeding 10,000 tokens.}
\label{fig:length}
\end{figure*}

\input{results_doc}

\subsection{Document Retrieval}
\label{exp:doc}

\paragraph{Dataset} 
The document retrieval task aims to rank document-length texts, which present the challenge of handling long input sequences~\cite{msmarco}.
As illustrated in Figure~\ref{fig:length}, the MS MARCO document ranking corpus has an average document length of around 1500 tokens.
Notably, only 24\% of the documents have fewer than 512 tokens, which is the maximum input length for most previous rerankers based on smaller pre-trained language models like BERT~\cite{bert}.

The standard solution to manage long sequences for retrieval is the MaxP strategy~\cite{maxp}, which involves dividing the document into overlapping segments and determining the document relevance score based on the segment with the highest score.
However, this process involves a heuristic pooling strategy and runs the risk of losing information spread across long contexts.
Recent language models pre-trained on longer sequences (e.g., 4096 tokens for LLaMA-2) offer the potential to represent document-length texts ``in one go'', reducing the need for segmentation.

By default we allow the retriever and reranker to take the first 2048 tokens as input without any segmentation, which is a reasonable trade-off between input sequence length and the cost of training.
This approach covers about 77\% of the documents in the corpus entirely.
We create the training data for the document retriever and reranker models based on the 300k training examples in the training set.
Similar to the approach in passage ranking, we sample the hard negative documents to train \RepLLaMA from the top-$100$ hard negatives from BM25 and our implementation of CoCondenser-MaxP.
Here, BM25 directly indexes the entire documents, while CoCondenser retrieves documents using the aforementioned MaxP strategy.
The hard negatives for \RankLLaMA are selected from the top-$100$ results of \RepLLaMA.

Evaluation of document retrieval is performed on the development split of the MS MARCO document ranking dataset, which contains 5193 queries.
Additionally, we evaluate our models on the TREC DL19/DL20 document ranking tasks, comprising 43 and 45 queries, respectively.

\paragraph{Implementation Details}
We follow a similar setup as in the passage ranking task to train both {\it document} \RepLLaMA and \RankLLaMA, with the same computing resources.
However, there are two key differences:
First, the models are trained with a batch size of 128, with each query sampling 7 hard negative passages.
Second, during inference, \RepLLaMA retrieves the top-$1000$ documents while \RankLLaMA reranks the top-$100$ documents that are retrieved by \RepLLaMA.
The document model also generates text embeddings with 4096 dimensions.
For the MS MARCO document corpus, this results in a 49G (flat) index after pre-encoding the entire corpus.

\paragraph{Results}
Table~\ref{tab:doc} reports the effectiveness of our \RepLLaMA--\RankLLaMA pipeline for full-document retrieval on the MS MARCO document corpus.
We see that both our retriever and reranker outperform existing methods.
\RepLLaMA achieves an MRR@100 score that is approximately 3 points higher than CoCondenser-MaxP, while \RankLLaMA exceeds (to our knowledge) the current state-of-the-art document reranker, MORES+~\cite{mores}, by 1 point in MRR@100.

We again emphasize that both our retriever and reranker do not require document segmentation and rank score aggregation.
Instead, \RepLLaMA directly consumes the entire document, and \RankLLaMA directly scores the relevance of the entire query--document pair.

\section{Ablation Study and Analysis}

\subsection{Full Fine-Tuning vs.\ LoRA}

\begin{table}[t]
\centering
\scalebox{0.8}{
\begin{tabular}{l|llll}
\toprule
     & Train & Dev  & DL19 & DL20 \\
\midrule
FT   & \bf 46.6  & \bf 41.6 & 72.8 & 69.9 \\
LoRA & 40.8  & 41.2 & \bf 74.3 & \bf 72.1 \\
\bottomrule
\end{tabular}}
\caption{Comparison of MRR@10 between full fine-tuning (FT) and LoRA when training \RepLLaMA for the passage retrieval task.}
\label{tab:lora}
\end{table}

When fine-tuning large language models, a key decision is whether to conduct full fine-tuning, which updates all parameters in the model, or to use a parameter-efficient method such as LoRA.
Table~\ref{tab:lora} compares the effectiveness of \RepLLaMA when trained with full fine-tuning and LoRA for the passage retrieval task.
Both models are trained on the training set for one epoch.

We see that full fine-tuning achieves an MRR@10 score that is approximately 6 points higher than with LoRA on the training set.
However, on the development set, full fine-tuning only improves effectiveness by 0.4 points compared to LoRA.
Interestingly, on the TREC DL19/DL20 datasets, which are derived from independent human judgments, LoRA demonstrates better effectiveness.
This suggests that full fine-tuning may be prone to overfitting on the training set distribution, while LoRA, with significantly fewer parameters, can generalizable better.
For this reason, all the models presented in our main experiments (Section~\ref{section:experiments}) use LoRA instead of full fine-tuning.

\subsection{Input Sequence Length}

As discussed in Section~\ref{exp:doc}, \RankLLaMA has the advantage of accommodating longer inputs compared to previous models like BERT since its LLaMA backbone was pre-trained with a longer context window.
We investigate the effects of varying the maximum training input length and inference input length on model effectiveness for the document reranking task.
Results presented in Figure~\ref{fig:len_abl} show a clear trend:\ the effectiveness of \RankLLaMA improves as the maximum training length increases from 512 to 2048, with the MRR@100 score improving from 48.5 to 50.3.
When the reranking input length is further increased to 4096, the MRR@100 score rises to 50.6.
This demonstrates the model's ability to exploit longer sequences for improved effectiveness.

\input{figure_length}

However, it is important to note that the gains plateau beyond a certain length, suggesting a point of diminishing returns.
The MRR@100 for the model trained with a length of 4096 is only 0.3 points higher than the model trained with a length of 2048, when evaluated on input lengths that match their training lengths.
Moreover, the model trained with a length of 4096 takes about 8 days to train using 16 $\times$ V100 GPUs, while the model with a length of 2048 takes about 4 days.
The same relative latency costs apply to inference as well.
Therefore, while \RankLLaMA can handle much longer input documents, it is crucial to balance this capability with the practical considerations of computational efficiency.

\section{Related Work}

\subsection{Large Language Models}
Pre-trained language models based on the Transformer architecture~\cite{transformers} have demonstrated impressive capabilities when fine-tuned for various downstream tasks since the advent of BERT~\cite{bert}.
Depending on their architecture, pre-trained Transformers can be classified into three categories:\ encoder-only models~\cite{bert, Liu2019RoBERTaAR, conneau-etal-2020-unsupervised}, encoder--decoder models~\cite{T5, lewis-etal-2020-bart}, and decoder-only models~\cite{radford_improving_2018}.
Decoder-only models like GPT/GPT-2 have been lauded for their simplicity in terms of model architecture and pre-training procedures~\cite{radford_improving_2018, Radford2019LanguageMA}.

Recent research has shown that scaling up LLMs by pre-training larger decoder-only models using larger and higher quality corpora can significantly enhance model capabilities for general-purpose NLP tasks such as question answering and code generation~\cite{Wei2022ChainOT, chen2021evaluating}.
This is achieved by fine-tuning the pre-trained LLMs with instruction-following data using reinforcement learning with human feedback.
Instruct\-GPT~\cite{instructGPT} and \mbox{GPT-4}~\cite{openai2023gpt4} are two popular representatives in this class of models.
Among the many implementations of open-source large language models, LLaMA~\cite{llama1, llama2} is among the most recent and among the top-performing on a variety of tasks.

\subsection{Multi-Stage Text Retrieval}
While multi-stage retrieval pipelines date back well over a decade~\cite{Matveeva_etal_SIGIR2006, Cambazoglu_etal_WSDM2010, Wang_etal_SIGIR2011}, they have benefited immensely from pre-trained language models such as BERT in recent years, starting with the monoBERT reranking model~\cite{monobert}.
\citet{Nogueira2019MultiStageDR} proposed a multi-stage retrieval pipeline that employs a BM25 retriever followed by two BERT-based reranking stages.
This design demonstrates the effectiveness of pre-trained language models in reranking tasks.
\RankLLaMA follows the same basic design as monoBERT.
The dense passage retriever (DPR) further proposed to fine-tune BERT to replace the BM25 retriever with a dense retrieval model in a bi-encoder design~\cite{karpukhin-etal-2020-dense}.
DPR encodes text into low-dimensional dense vector representations and treats retrieval as a nearest-neighbor search task.
\RepLLaMA follows the same bi-encoder design.

Several solutions have been introduced to enhance the effectiveness of retrievers and rerankers in a multi-stage pipeline.
On the retriever side, works such as ANCE~\cite{ance}, Rocket\-QA~\cite{qu-etal-2021-rocketqa}, CoCondenser~\cite{gao-callan-2022-unsupervised}, RetroMAE~\cite{xiao-etal-2022-retromae}, and SimLM~\cite{wang-etal-2023-simlm}, have shown that augmenting the training data with hard negative mining or continuous retrieval-oriented pre-training can improve the effectiveness of dense retrievers.
On the reranker side, monoT5~\cite{nogueira-etal-2020-document} and monoELECTRA~\cite{monoElectra} demonstrated that initializing the reranker with a custom pre-trained model can enhance effectiveness.
\citealp{Gao2021RethinkTO} proposed using a contrastive loss for reranker training to replace the default pairwise loss.
\citet{rankT5} studied the use of T5 as a reranker, analyzing the influence of different model architectures and loss functions.
However, directly fine-tuning modern billion-parameter-scale large language models for multi-stage retrieval has not been explored to date.

Recently, LLMs have demonstrated impressive effectiveness when prompted to perform few-shot or zero-shot text generation.
As mentioned in the introduction, researchers have cast reranking as text generation.
These models can be leveraged to directly generate a reordered list of candidates, e.g., LRL~\cite{LRL}, RankGPT~\cite{RankGPT}, RankVicuna~\cite{pradeep2023rankvicuna}.
Alternatively, they can compare passages in a pairwise manner, e.g., PRP~\cite{PRP}.
Although prompt-based methods have shown good zero-shot effectiveness, they require multiple decoding passes, thus making them slow and non-parallelizable.
Furthermore, reranking with prompts makes it difficult to exploit available human judgments such as MS MARCO~\cite{msmarco} to further improve effectiveness.
Finally, these approaches do not allow for joint reranker--retriever optimization.
In contrast, we address all these issues.

Our work is most similar to GPT-XXL~\cite{ni-etal-2022-large} and SGPT~\cite{muennighoff2022sgpt}, which also used billion-parameter-scale models as backbones of dense retrievers, achieving better zero-shot effectiveness than smaller models.
However, LLaMA has demonstrated even better effectiveness on natural language generation tasks, suggesting that it might serve as a better backbone and warranting further exploration.
The cpt-text model~\cite{neelakantan2022text}, initialized with the 175-billion-parameter GPT-3 model, also shows strong zero-shot effectiveness.
However, cpt-text is not an open-source model.
Additionally, none of the models referenced above are fully optimized for a multi-stage retrieval pipeline.
Our \RepLLaMA and \RankLLaMA models are fully open-source and optimized for multi-stage retrieval, achieving state-of-the-art effectiveness on both retrieval and reranking, for both in-domain and out-of-domain evaluations.

\section{Conclusion}

The successful application of large language models in generative tasks has sparked interest in their potential to enhance retrieval.
In this study, we demonstrate that it is possible to fine-tune a large model to act as a dense retriever (\RepLLaMA) and a pointwise reranker (\RankLLaMA), thereby establishing an effective, state-of-the-art multi-stage retrieval system that outperforms smaller models built on the same basic design.
Moreover, our approach offers greater optimization and efficient inference potential than recent methods that prompt large language models for text reranking in a generative manner.
This work underscores the potential of leveraging LLMs for retrieval tasks in the future, which we continue to explore.

\section*{Acknowledgments}

This research was supported in part by the Natural Sciences and Engineering Research Council (NSERC) of Canada.


\balance

\bibliography{anthology,custom}
\bibliographystyle{acl_natbib}

\end{document}

%% file: results_passage.tex
\begin{table*}[t]
\centering
\scalebox{0.79}{
\begin{tabular}{l|c|cr|cc|c|c}
\toprule
& \textbf{Model} & \multicolumn{2}{c|}{\textbf{Source}} & \multicolumn{2}{c|}{\bf DEV} & \bf DL19   & \bf DL20  \\
&  \textbf{size} & \textbf{prev.} & \textbf{top-}$k$ & \textbf{MRR@10} & \textbf{R@1k} & \textbf{nDCG@10} & \textbf{nDCG@10} \\
\midrule
\multicolumn{8}{c}{\textit{Retrieval}} \\
BM25~\cite{Lin_etal_SIGIR2021_Pyserini} & -    &   -   & $|C|$ & 18.4  & 85.3 & 50.6 & 48.0 \\
ANCE~\cite{ance}         & 125M &   -   & $|C|$ & 33.0 & 95.9 & 64.5 & 64.6 \\
CoCondenser~\cite{gao-callan-2022-unsupervised}  & 110M &   -   & $|C|$ & 38.2 & 98.4 & 71.7 & 68.4 \\
GTR-base~\cite{ni-etal-2022-large}     & 110M &   -   & $|C|$ & 36.6 & 98.3 &   -  &  -   \\
GTR-XXL~\cite{ni-etal-2022-large}      & 4.8B &   -   & $|C|$ & 38.8 & 99.0 &   -  &  -   \\
OpenAI Ada2~\cite{neelakantan2022text} & ? & - & $|C|$ & 34.4 & 98.6 & 70.4 & 67.6 \\
bi-SimLM~\cite{wang-etal-2023-simlm}   & 110M &   -   & $|C|$ &  39.1 & 98.6 & 69.8 & 69.2 \\
\RepLLaMA     & 7B  &   -   & $|C|$ & \bf 41.2 & \bf 99.4 & \bf 74.3 & \bf 72.1 \\
\midrule
\multicolumn{8}{c}{\textit{Reranking}} \\
monoBERT~\cite{Nogueira2019MultiStageDR}     & 110M & BM25     & 1000  & 37.2 & 85.3 & 72.3 & 72.2 \\
cross-SimLM~\cite{wang-etal-2023-simlm}  & 110M & bi-SimLM & 200   & 43.7 & 98.7  &   74.6  &   72.7  \\
RankT5~\cite{rankT5}       & 220M & GTR      & 1000  & 43.4 & 98.3  &   -  &   -  \\
\RankLLaMA    & 7B   & \RepLLaMA & 200   & 44.9 & 99.4   &  75.6   &   77.4  \\
\RankLLaMA-13B    & 13B  & \RepLLaMA & 200   & \bf 45.2 &  \bf 99.4   &  \bf 76.0   &   \bf 77.9  \\
\midrule
RankVicuna~\cite{pradeep2023rankvicuna} & 7B & BM25 & 100 & - & - & 66.8 & 65.5 \\
PRP~\cite{PRP} & 20B & BM25 & 100 & - & - & 72.7 & 70.5 \\
RankGPT$_{3.5}$~\cite{RankGPT} & ? & BM25 & 100 & - & - & 65.8 & 72.9\\
RankGPT$_4$~\cite{RankGPT}   & ? & RankGPT$_{3.5}$ & 30 & - & - & 75.6 & 70.6 \\
\bottomrule
\end{tabular}
}
\caption{The effectiveness of \RepLLaMA and \RankLLaMA on the MS MARCO passage corpus compared to existing methods.
For the retriever, we compare against models trained with binary human judgments, without distillation from a reranker.
Evaluation figures are copied from the original papers except for OpenAI Ada2, which is the successor to cpt-text~\cite{neelakantan2022text} and available as a commercial API. The effectiveness numbers of Ada2 are taken from~\citet{Lin_etal_arXiv2023_OpenAI}.
}
\label{tab:passage}
\end{table*}

%% file: results_beir.tex
\begin{table*}[t]
\centering
\scalebox{0.775}{
\begin{tabular}{l|cccccc|cccc}
\toprule
 & BM25 & GTR-XXL & cpt-text-XL & Ada2 & SGPT & \RepLLaMA & RankT5 & \RankLLaMA  & \RankLLaMA   \\
 model size & - & 4.8B & 175B & ? & 5.8B & 7B  & 220M  & 7B & 13B \\
 add. pretrain & - & Y & Y & ? & Y & N & - & - & - \\
 \midrule
Arguana       & 39.7 & 54.0 & 43.5 & 56.7 & 51.4 & 48.6 & 33.0 & 56.0 & 50.8 \\
Climate-FEVER & 16.5 & 26.7 & 22.3 & 23.7 & 30.5 & 31.0 & 21.5 & 28.0 & 29.2 \\
DBPedia       & 31.8 & 40.8 & 43.2 & 40.2 & 39.9 & 43.7 & 44.2 & 48.3 & 48.7 \\
FEVER         & 65.1 & 74.0 & 77.5 & 77.3 & 78.3 & 83.4 & 83.2 & 83.9 & 86.2 \\
FiQA          & 23.6 & 46.7 & 51.2 & 41.1 & 37.2 & 45.8 & 44.5 & 46.5 & 48.1 \\
HotpotQA      & 63.3 & 59.9 & 68.8 & 65.4 & 59.3 & 68.5 & 71.0 & 75.3 & 76.4 \\
NFCorpus      & 32.2 & 34.2 & 40.7 & 35.8 & 36.2 & 37.8 & 38.1 & 30.3 & 28.4 \\
NQ            & 30.6 & 56.8 &  -   & 48.2 & 52.4 & 62.4 & 61.4 & 66.3 & 66.7 \\
Quora         & 78.9 & 89.2 & 63.8 & 87.6 & 84.6 & 86.8 & 83.1 & 85.0 & 81.7 \\
SCIDOCS       & 14.9 & 16.1 &  -   & 18.6 & 19.7 & 18.1 & 18.1 & 17.8 & 19.1 \\
SciFact       & 67.9 & 66.2 & 75.4 & 73.6 & 74.7 & 75.6 & 75.0 & 73.2 & 73.0 \\
TREC-COVID    & 59.5 & 50.1 & 64.9 & 81.3 & 87.3 & 84.7 & 80.7 & 85.2 & 86.1 \\
Touche-2020   & 44.2 & 25.6 & 29.1 & 28.0 & 25.4 & 30.5 & 44.0 & 40.1 & 40.6 \\
\midrule
Average       & 43.7 & 49.3 & - & 52.1 & 52.1 & \bf55.1 & 53.7 & \bf56.6 & 56.5  \\
\bottomrule
\end{tabular}
}
\vspace{0.2cm}
\caption{Zero-shot effectiveness of \RepLLaMA and \RankLLaMA on BEIR datasets. The ``\textit{add.\ pretrain}'' row indicates whether the retriever model has undergone additional contrastive pre-training before supervised fine-tuning.
The zero-shot effectiveness numbers of Ada2 are taken from~\citet{kamalloo-etal-2023-evaluating-embedding}.
}
\label{tab:beir}
\end{table*}

%% file: results_doc.tex
\begin{table*}[t]
\centering
\scalebox{0.78}{
\begin{tabular}{l|c|cr|c|cc|c|c}
\toprule
& \textbf{Model} & \multicolumn{2}{c|}{\textbf{Source}} & \textbf{Seg.} & \multicolumn{2}{c|}{\bf Dev} & \bf DL19   & \bf DL20  \\
&  \textbf{size} & \textbf{prev.} & \textbf{top-}$k$ & \textbf{Y/N} & MRR@100 & R@1k & nDCG@10 & nDCG@10 \\
\midrule
\multicolumn{8}{c}{\textit{Retrieval}} \\
BM25~\cite{Lin_etal_SIGIR2021_Pyserini}  & -    &   -   & $|C|$ & N & 23.0  & 85.3 & 51.8 & 52.9 \\
BM25-Q2D~\cite{Pradeep2021TheED}             & -    &   -   & $|C|$ & Y & 31.8  & 94.9 & 61.2 & 59.6 \\
CoCondenser-MaxP & 110M &   -   & $|C|$ & Y  &42.5  & 93.9    & 64.8 & \bf 64.0 \\
\RepLLaMA        & 7B   &   -   & $|C|$ & N  & \bf 45.6 & \bf 98.9 & \bf 65.0 & 63.2 \\
\midrule
\multicolumn{8}{c}{\textit{Reranking}} \\
monoT5~\cite{Pradeep2021TheED}  & 3B   & BM25-Q2D  & 10000 & Y & 41.1 & 94.9  &  -  & 
 -  \\
MORES+~\cite{mores}& 110M & CoCondenser    & 100   & Y & 49.3 & -      &    -    & -  \\
\RankLLaMA    & 7B   & \RepLLaMA       & 100   & N & \bf 50.3 & \bf 98.9 & \bf 67.7 & \bf 67.4 \\

\bottomrule
\end{tabular}
}
\caption{The effectiveness of \RepLLaMA and \RankLLaMA on the MS MARCO document corpus compared to existing methods.}
\label{tab:doc}
\end{table*}

%% file: figure_length.tex
\begin{figure}[t]
\centering
\includegraphics[width=0.45\textwidth]{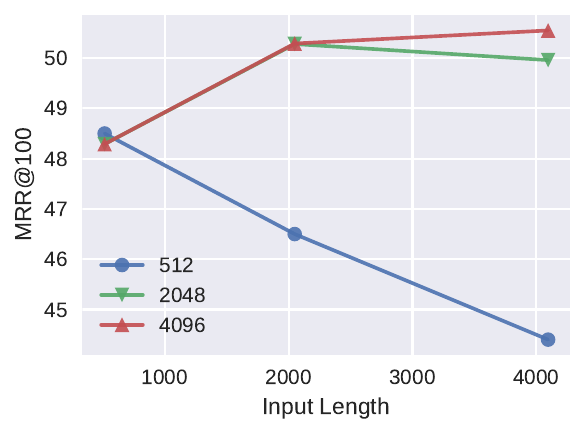}
\caption{Comparison of document ranking MRR@100 scores for \RankLLaMA trained with different maximum input lengths and evaluated using different maximum input lengths.
Each line represents a model trained with a specific maximum length, while points along the line indicate the effectiveness when varying the input length during inference (reranking).}
\label{fig:len_abl}
\end{figure}